\documentclass[twocolumn,prb,aps,showpacs,superscriptaddress]{revtex4-1}
\usepackage{graphicx}
\usepackage{amssymb}
\usepackage{textcomp}
\def\be{\begin{equation}}
\def\ee{\end{equation}}
\def\bea{\begin{eqnarray}}
\def\eea{\end{eqnarray}}
\def\bdm{\begin{displaymath}}
\def\edm{\end{displaymath}}
\def\ba{\begin{array}}
\def\ea{\end{array}}

\begin{document}

\title{Probing Layer Localization in Twisted Graphene Bilayers via Cyclotron Resonance }

\author{Chi-Ken Lu}
\author{H.~A. Fertig}
\affiliation{Department of Physics, Indiana University, Bloomington, Indiana 47405, USA}
%\date{\today }

\begin{abstract}

Electron wavefunctions in twisted bilayer graphene may have a strong single layer character
or be intrinsically delocalized between layers, with their nature often determined by
how energetically close they are to the Dirac point.
In this paper, we demonstrate that in magnetic fields, optical absorption (cyclotron resonance)
spectra contain signatures which may be used to distinguish the nature of these
wavefunctions at low energies, as well as to locate low energy critical
points in the zero-field energy spectrum. Optical absorption for two different configurations
-- electric field parallel and perpendicular to the bilayer --
are calculated, which are shown to have different selection rules with respect
to which states are connected by the perturbation.
Interlayer bias further distinguishes transitions involving states of a single
layer nature from those with support in both layers. For doped systems,
a sharp increase in intra-Landau level absorption occurs with increasing field as
the level passes
through the zero-field saddle point energy, where the states change character from
single layer to bilayer.

\end{abstract}


\maketitle

\section{Introduction}

Bilayer graphene~\cite{McCann} is a remarkable electronic material because it supports
a variety of spectra, determined by the relative angle of the nearest neighbor bonds
of the two layers~\cite{tBLG1,tBLG_EAndrea}, which may be continuously tuned
with interlayer bias~\cite{GapBilayer}.  The most common configuration, Bernal ("AB")
stacking,
can be turned into a band insulator by application of an interlayer bias, and this
behavior allows the generation of domain walls with topologically non-trivial electronic
properties, both in zero~\cite{IMartin,Jacoby,SanJose}
and in finite~\cite{huang,mazo} magnetic fields.  At the other extreme, AA-stacked graphene
bilayers~\cite{liu} can be created and should support interesting electronic
phases~\cite{rakhmanov,brey_AA}.  Interpolating between these two extremes is
twisted bilayer graphene (tBLG), in which a periodic Moir\'e pattern with regions
of AB and AA stacking are contained in each unit cell~\cite{tBLG1,brey_2011}.
In momentum space, an important feature of this structure is the existence of low
energy saddle points, which lead to observable van Hove singularities in the
density of states \cite{vanHove,tBLG_EAndrea,Lee11a,Kim12,Ohta}. Interactions in
the setting of
such an enhanced density of states could lead to
unusual electronic states \cite{SCGraphene,natPhysSC,SCtBLG}.
Although van Hove singularities are present in the density of states for single
layer graphene, placing the Fermi level in their vicinity requires a high level
of doping which has so far not been achievable.  From this perspective,
tBLG offers an advantage because the van Hove singularity is accessible at
a much lower doping level. It is therefore important to understand the
effects of the saddle points which underpin the van Hove singularity on the
electronic structure of tBLG, and how they might be probed experimentally.
In this paper we study this question for the non-interacting system,
focusing on the effects of a magnetic field, and in particular how cyclotron
resonance can uncover various properties of the spectrum and electronic
wavefunctions.

The low-energy dynamics of electrons in tBLG is governed by Dirac cones
from the top and bottom layers, and a periodic vertical hopping between them~\cite{Mele}.
The locations of the Dirac points in momentum space are different for each of the two layers
because of the twist angle.
Near the neutrality point (zero energy), the Dirac dispersions remain linear in spite
of the interlayer coupling, although
the Fermi velocity is reduced~\cite{tBLG1}.  Away from this energy, near the value at which there are states
at the same energy and momentum from both layers in the absence of interlayer coupling,
tunneling has strong effects:
the bands split into an upper and lower branch, the lower one containing a
saddle point and the upper a quadratic minimum \cite{LuFertig14}. In terms of the density of states, the former leads to a logarithmic divergence while the latter is revealed by a sharp jump~\cite{Louis,Moon2}.
In the presence of a magnetic field,
below the saddle point energy one finds Landau levels which
are similar to those of single-layer graphene but with a two-fold
degeneracy due to the layer degree of freedom~\cite{deGail1,Choi}.
Above the saddle point, however, tunneling between the layers becomes
qualitatively relevant. The energy interval between the saddle point and
the quadratic minimum represents a transition region that supports interweaving
electron-like and hole-like energy levels which can be understood within
a semiclassical framework~\cite{LuFertig14}.  In particular, one may to an excellent
approximation regard the semiclassical orbits as residing in one or the other layer
for energies below the saddle point.  Above it, the orbits necessarily periodically tunnel
between the layers~\cite{LuFertig14}.
Above the energy of the quadratic minimum, the semiclassical orbits become more complicated, and
the full intricacy of bands and gaps that characterizes the ``Hofstadter spectrum'' becomes apparent~\cite{Hofstadter,Moon1,Hofstadter_exp1,Hofstadter_exp2,Bistritzer2}.

In what follows we explore what is revealed about the crossover behavior in the vicinity
of the saddle point in cyclotron resonance \cite{Jiang07,Deacon07,Henriksen08a,Orlita12}.
A unique aspect of a bilayer system is the possibility to orient the electric field polarization
of the incoming radiation perpendicular to the layers.  Such radiation cannot be absorbed
in single layer systems, but in bilayers it can be, provided the wavefunctions of the
initial and/or final states involve both layers.
Thus a comparison of absorption spectra for parallel
(i.e., in-plane) and perpendicular orientations can help distinguish regions of the
spectrum supporting different types of wavefunctions.
Our analysis employs a continuum model of tBLG \cite{tBLG1},
which is used to describe a circular system of finite size.  Within this
model, with an appropriate gauge choice
the model Hamiltonian possesses three-fold rotational symmetry, allowing the eigenstates to be classified into three sectors. Interestingly, incoming radiation polarized perpendicular
to the layers induces only intra-sector excitations, while parallel polarization induces
only inter-sector excitations.

Overall, our calculations demonstrate that the absorption spectrum for tBLG is very
different than that of two decoupled layers.  In particular there is very strong
absorption for transitions involving states below the negative energy and/or above
the positive energy quadratic band, such that the overall absorption is greatly
enhanced over that of two single layers, as we show in more detail below.
As might be expected, density of states calculations show that
states primarily localized in a single
layer can be distinguished from ones distributed between two layers by their behavior
with interlayer bias: the energies of the former are sensitive to bias while the
latter are not \cite{comment}.  Moreover, we find that Landau levels become highly
broadened as they pass through the saddle point energy with increasing
magnetic field.  In cyclotron resonance, we find that peaks associated with intralayer initial
and final states are insensitive to interlayer bias, while transitions involving interlayer
states shift noticeably with bias.  The energy scale for which states above the quadratic
band edge become involved and absorption starts to increase markedly is also sensitive to this bias.
Finally, we show that the large broadening of a Landau level when it is in the
vicinity of the saddle point greatly enhances intra-Landau level absorption.
In principle this effect could be used to locate the energy of the saddle point
in the zero field spectrum.

This article is organized as follows.  In Section II we describe the model
Hamiltonian used in our analysis, and how the three-fold symmetry is
exploited in the calculation.  Section III is devoted to density of states
calculations, which are compared and contrasted with the expected
results for single layers.  Section IV contains results of calculated cyclotron
resonance spectra in various circumstances.  We conclude with a discussion
and summary in Section V.

\section{Model Hamiltonian and three-fold symmetry}

Twisted bilayer graphene with an interlayer bias $u$ can be modeled by a continuum Hamiltonian of the form \cite{tBLG1,Bistritzer2,SanJose}

\be
 H=\left(\begin{array}{cccc}
    \mathcal H_D & \mathcal V\\%w\sum_{i=0,1,2}V_i   \\
    %w\sum_{i=0,1,2}V_i^{\dag}
    \mathcal V^{\dag}& \mathcal H_D\end{array}\right)
    +
    \frac{u}{2}\left(\begin{array}{cccc}
     \mathbb{I}_2& 0\\%w\sum_{i=0,1,2}V_i   \\
    %w\sum_{i=0,1,2}V_i^{\dag}
    0 &  -\mathbb{I}_2\end{array}\right)\:,\label{BasicModel}
\ee
where $\mathcal H_D=v_F(\hat\sigma_x {p}_1+\hat\sigma_y{p}_2)$ is the single-layer graphene
Dirac Hamiltonian. In this expression $\sigma_{x,y}$ and $\mathbb{I}_2$ are the Pauli matrices and the identity matrix, respectively, which act on the sublattice space. The second term in
Eq.\ (\ref{BasicModel}) implements the interlayer bias $u$. The off-diagonal contribution $\mathcal V=w\sum_{i=0,1,2}e^{-i{\bf q_i}\cdot{\bf r}}\hat V_i$ is a sum of three interlayer hopping matrices,

\be
	\hat V_0=\left(\begin{array}{cccc}
    	1 & 1   \\
    	1 & 1
    	\end{array}\right)\:,
  	\ \hat V_1=\left(\begin{array}{cccc}
    	\bar z & 1   \\
    	z &  \bar z
    	\end{array}\right)\:,
  	\ \hat V_2=\left(\begin{array}{cccc}
    	 z & 1   \\
    	\bar z & z
    	\end{array}\right)\:,\label{three_matrices}
\ee with the complex number $z \equiv \exp(i2\pi/3)$ and $\bar z$ is its complex conjugate. Within the Brillouin zone of the Moir\'e pattern, the three wave vectors ${\bf q_0}=k_{\theta}(0,-1)$, ${\bf q_1}=k_{\theta}(-\sqrt{3}/2,1/2)$, and ${\bf q_2}=k_{\theta}(\sqrt{3}/2,1/2)$ join a Dirac point of one layer with a corresponding Dirac point of the other. Because of the twist, the Dirac points of opposite layers are separated by $k_{\theta}=2k_D\sin\frac{\theta}{2}$ with $\theta$ the rotational angle \cite{tBLG1},
and
$k_D$ the distance from the $\Gamma$ point to the Dirac point in the Brillouin zone
of single layer graphene.

In the presence of a uniform magnetic field along the $z$ axis, the momentum operators
are replaced by the covariant momentum operators, via $\pi_{1,2} \rightarrow p_{1,2}-A_{1,2}$,
in the Hamitonian $\mathcal H_D$. In this paper, we use the circular gauge ${\bf A}=B/2(-y,x)$ in order to exploit the rotational symmetry of the system. The eigenstates of $\mathcal H_D$ are

\be
	\psi_{n,m,\pm}=\left(\begin{array}{c}|n,m\rangle\\\pm|n-1,m+1\rangle\end{array}\right)\:,
\ee with the basis ket states given by

\be
	|n,m\rangle=\frac{(a^{\dag})^n}{\sqrt{n!}}\frac{(b^{\dag})^{n+m}}{\sqrt{(n+m)!}}|0,0\rangle\:,
\ee with $n$ and $m$ the Landau index and angular momentum, respectively \cite{Jain}, and $a^{\dag}$ ($b^{\dag}$) the raising operator for the Landau level index (angular momentum). The angular
momentum of these states is captured by the relation $L_z|n,m\rangle=m|n,m\rangle$, with $L_z=xp_y-yp_x$. The Hilbert space for this problem is defined by integers $n\geq 0$ and $m\geq -n$.
In practice, we introduce numerical cutoffs for the upper values of $n$ and $m$, which are
large enough that the results presented below are insensitive to their precise values.
Note that in introducing these cutoffs, we are effectively considering a finite size
system \cite{Jain}, but the portions of the resulting spectra that we analyze involve wavefunctions
well-away from the edge.

For the single layer Dirac system,
the rotation operator about the $z$ axis,

\be
	U({\phi})=e^{-i\phi L_z}e^{-i\frac{\phi}{2}\sigma_z}\:,
\ee can be shown to commute with $\mathcal H_D=\vec \pi\cdot\vec\sigma$ for arbitrary angle $\phi$, and one can easily verify that $U(\phi)\psi_{n,m}=e^{-i(m+1/2)\phi}\psi_{n,m}$.
The full $U(1)$ symmetry of such rotations is broken down to a $C_3$
rotational symmetry
by the
interlayer coupling $\mathcal V$. This can be seen by first noting that $U^{\dag}(\frac{2\pi}{3})e^{-i{\bf q}_i\cdot{\bf r}}U(\frac{2\pi}{3})= e^{-i{\bf q}_i\cdot\mathcal R{\bf r}}=e^{-i({\mathcal R}^{-1}{\bf q}_i)\cdot{\bf r}}$.
For the three wavevectors ${\bf q}_i$, the rotation operator
$\mathcal R^{-1}$ acts cyclically, rotating ${\bf q}_{0,1,2}$ into ${\bf q}_{1,2,0}$.
From this, one may show that the vertical hopping matrices follow an analogous cyclic relation, $U^{\dag}\hat V_{0,1,2}U=z\hat V_{1,2,0}$. Consequently, the interlayer hopping matrix under a three-fold rotation transforms as $\mathcal V\mapsto z\mathcal V$. We may absorb the additional factor $z$ by defining the full rotational operator

\be
	G=\bar \omega\exp[i\frac{\pi}{3}\tau_z]\otimes U(\frac{2\pi}{3})\:,\label{G_sym}
\ee where the Pauli matrix $\tau_3$ acts on the layer index and the complex number $\omega=\exp(i\pi/3)$. This symmetry operator commutes with the full Hamiltonian, $H$ in Eq.\ (\ref{BasicModel}).

Thus, eigenstates of $H$ may simultaneously be eigenstates of $G$.
A general state in the Hilbert space can always be written in the form

\be
	\Phi=\sum c^{n_1,n_2,n_3,n_4}_{m_1,m_2,m_3,m_4}\left(\begin{array}{c} |n_1,m_1\rangle\\
	|n_2,m_2\rangle\\ |n_3,m_3\rangle \\ |n_4,m_4\rangle\end{array}\right)\:,\label{Phi}
\ee
where the summation is over the Landau indices $n$ and the angular momenta $m$.
We wish to find conditions on the expansion coefficients which will make a particular
state an eigenstate of $G$.  Writing such states as $\Phi_{\nu}$, with
$\nu\in\mathbb Z_3=\{0,1,2\}$ determining the eigenvalue of $G$ by the relation

\be
	G\Phi_{\nu}=z^{\nu}\Phi_{\nu}\:,
\ee
one finds that the non-zero coefficients obey relations among their angular momentum indices
of the form
\be
	m_1\sim m_4\sim \nu\:,\ m_2\sim (m_1+1)\:,\ m_4\sim (m_3+1)\:.\label{Phi_cond}
\ee
In the above equations
the relation $\sim$ is defined for a pair of integers $(p,q)$ such that $p\sim q$ if $3\bigm|(p-q)$
(i.e., $p-q$ is evenly divisible by 3.)
Consequently, the operator $G$ divides the Hilbert space into {\it three} orthogonal sectors labelled by $\nu$, and the vanishing commutator of $G$ and $H$ guarantee $\langle\Phi_{\mu}|H|\Phi_{\nu}\rangle\propto\delta_{\mu\nu}$.
Thus, eigenvalues and eigenstates of the system can be found separately for each sub-Hilbert space
defined by the index $\nu$.  In practice, this reduces the size of matrices one needs
to diagonalize $H$ for given maximum values of $m$ and $n$ by approximately a factor
of three.

The matrix elements of $H$ involve sub-matrix elements of the operators $\mathcal H_D$
and $\mathcal V$.  The former can easily be computed using the algebra of the operators
$a$ and $b$.
The matrix elements of $\mathcal V$ require the integrals $\mathcal F^{n_1m_1}_{n_2m_2}({\bf q}_i)\equiv\langle n_1,m_1| e^{i{\bf q}_i\cdot{\bf r}}|n_2,m_2\rangle$,
which may be computed following methods detailed in Ref.\ \onlinecite{Jain}, with the result

\bea
  \mathcal F^{n_1m_1}_{n_2m_2}=e^{i(m_2-m_1)\theta_i}(i|\alpha|)^{|n_1-n_2|+|n_1-n_2+m_1-m_2|}\\
  e^{-|\alpha|^2}\sqrt{\frac{{\rm min}(n_1,n_2){\rm min}(n_1+m_1,n_2+m_2)}{{\rm Max}(n_1,n_2){\rm Max}(n_1+m_1,n_2+m_2)}}\nonumber\\
  \mathcal L^{|n_1-n_2|}_{{\rm{min}}(n_1,n_2)}(|\alpha|^2)\mathcal L^{|n_1-n_2+m_1-m_2|}_{{\rm min}(n_1+m_1,n_2+m_2)}(|\alpha|^2)\nonumber\:,
\eea where the parameter $\alpha=\frac{i\ell}{\sqrt{2}}(q_x+iq_y)$ and the symbol $\mathcal L$ stands for the associated Laguerre polynomial. $\theta_i$ refers to the angle between the vector ${\bf q}_i$ and the $x$ axis.

Finally, we note that
a perturbing electric field such as one encounters in cyclotron resonance
may be polarized along the
$\hat z$ axis.  This does not break the three-fold rotational symmetry,
so only transitions among states with the same value of $\nu$ will be induced.
On the other hand, if the perturbing field is in the plane, the three-fold symmetry is broken,
in such a way that only transitions that change $\nu$ can have non-zero weight, as
demonstrated below. Before turning to this, we discuss the energy spectra obtained for the unperturbed system.

\section{Landau levels in circular gauge}

\begin{widetext}

\begin{figure}
\input{epsf}
\includegraphics[width=0.4\textwidth]{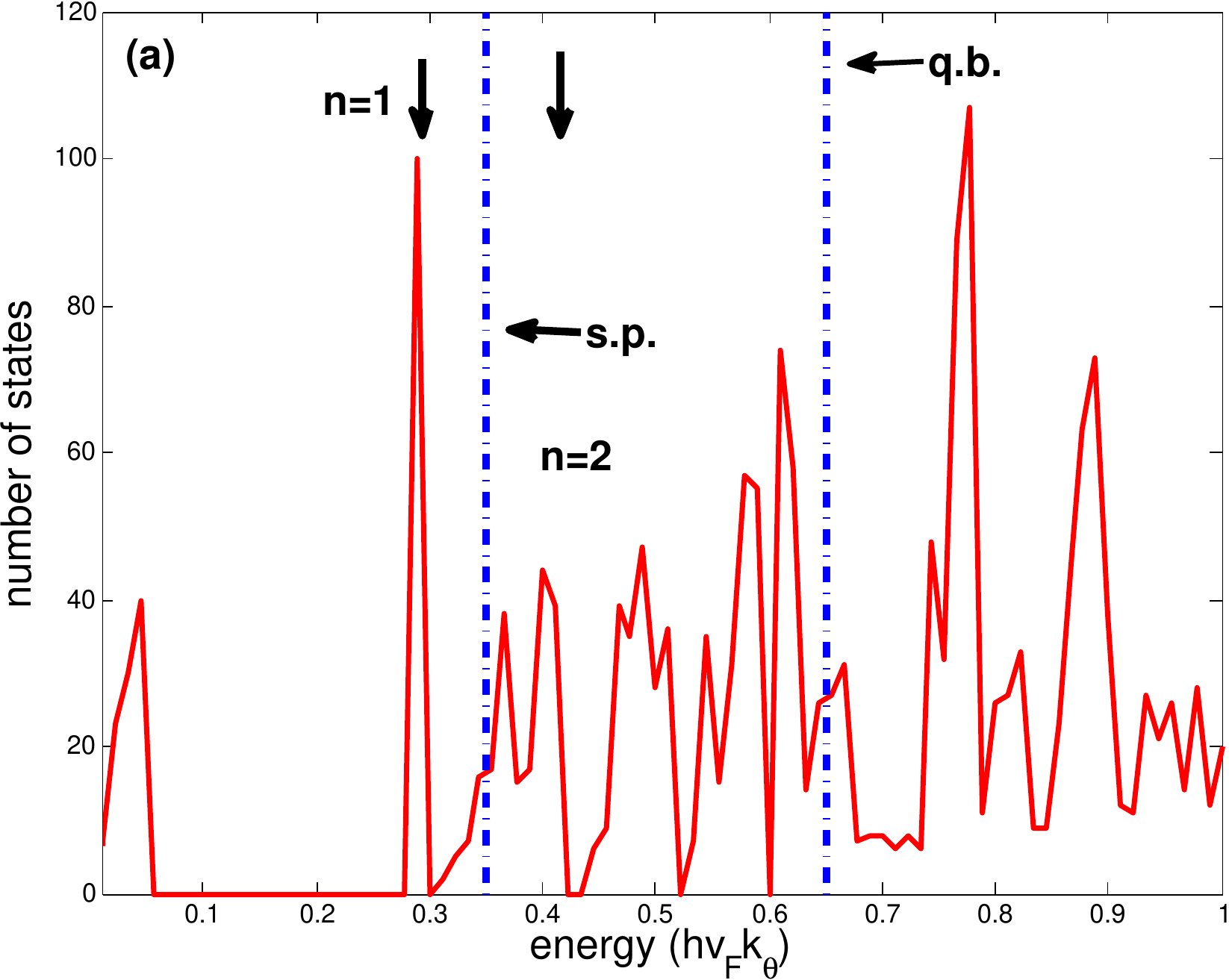}
\includegraphics[width=0.4\textwidth]{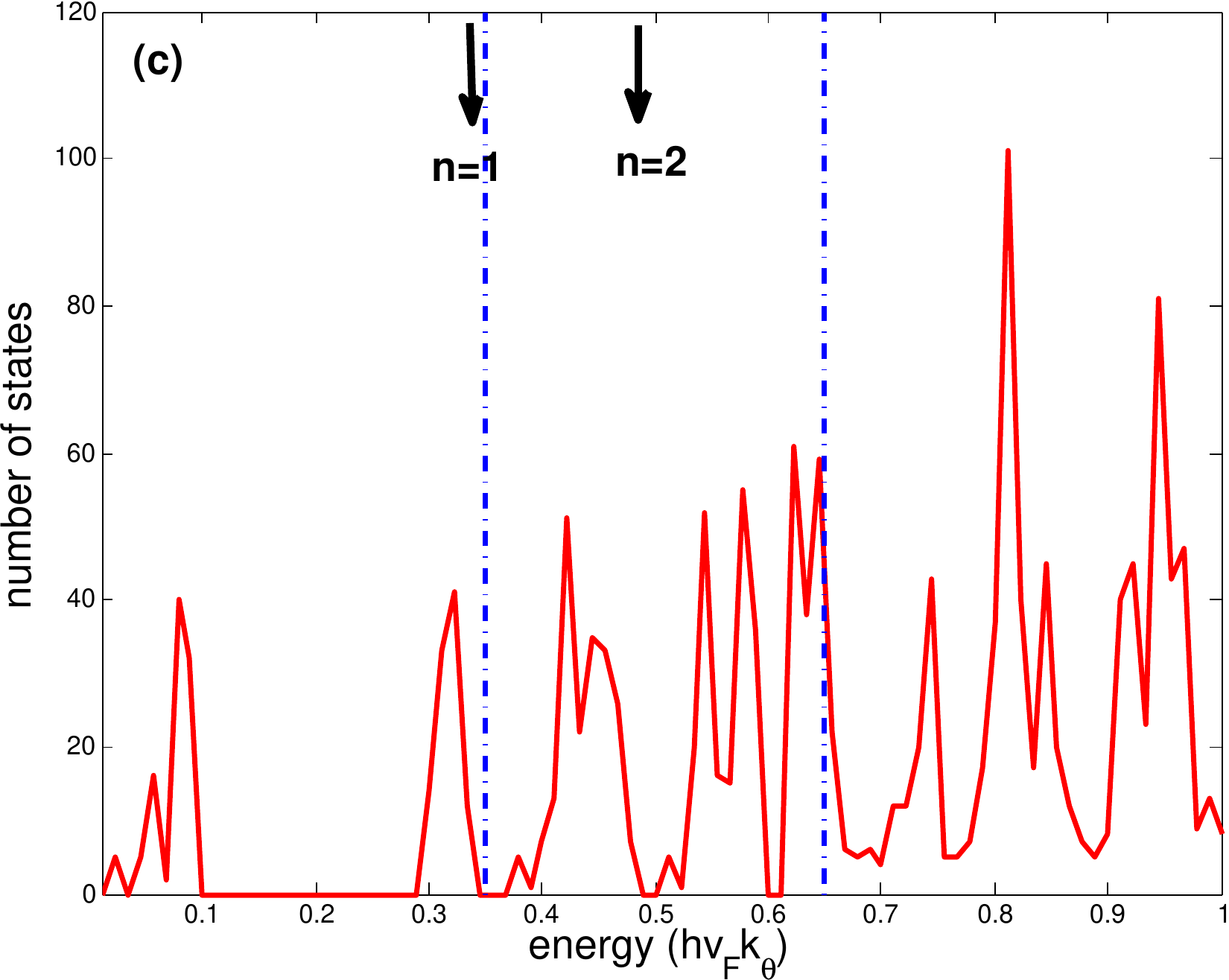}
\includegraphics[width=0.4\textwidth]{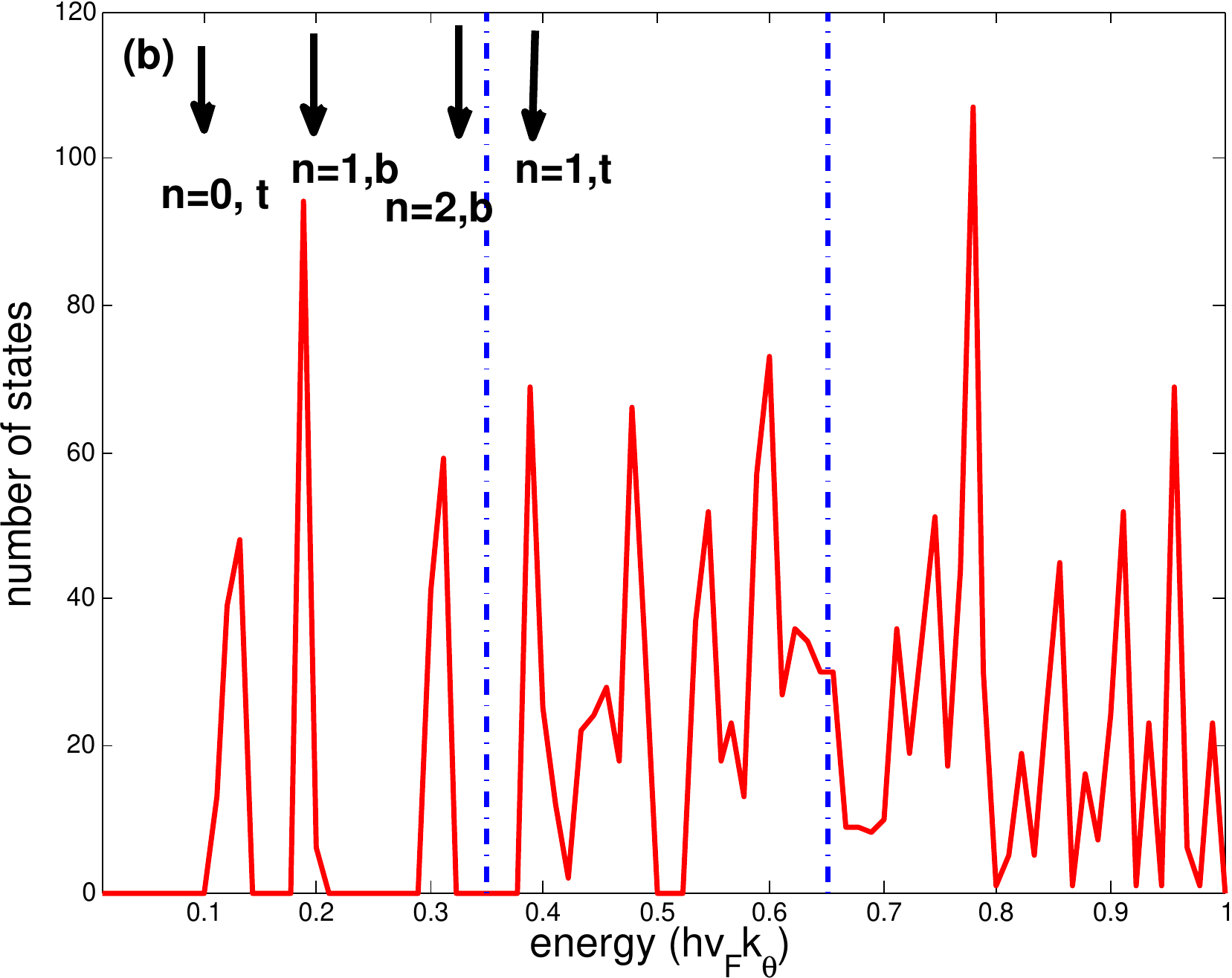}
\includegraphics[width=0.4\textwidth]{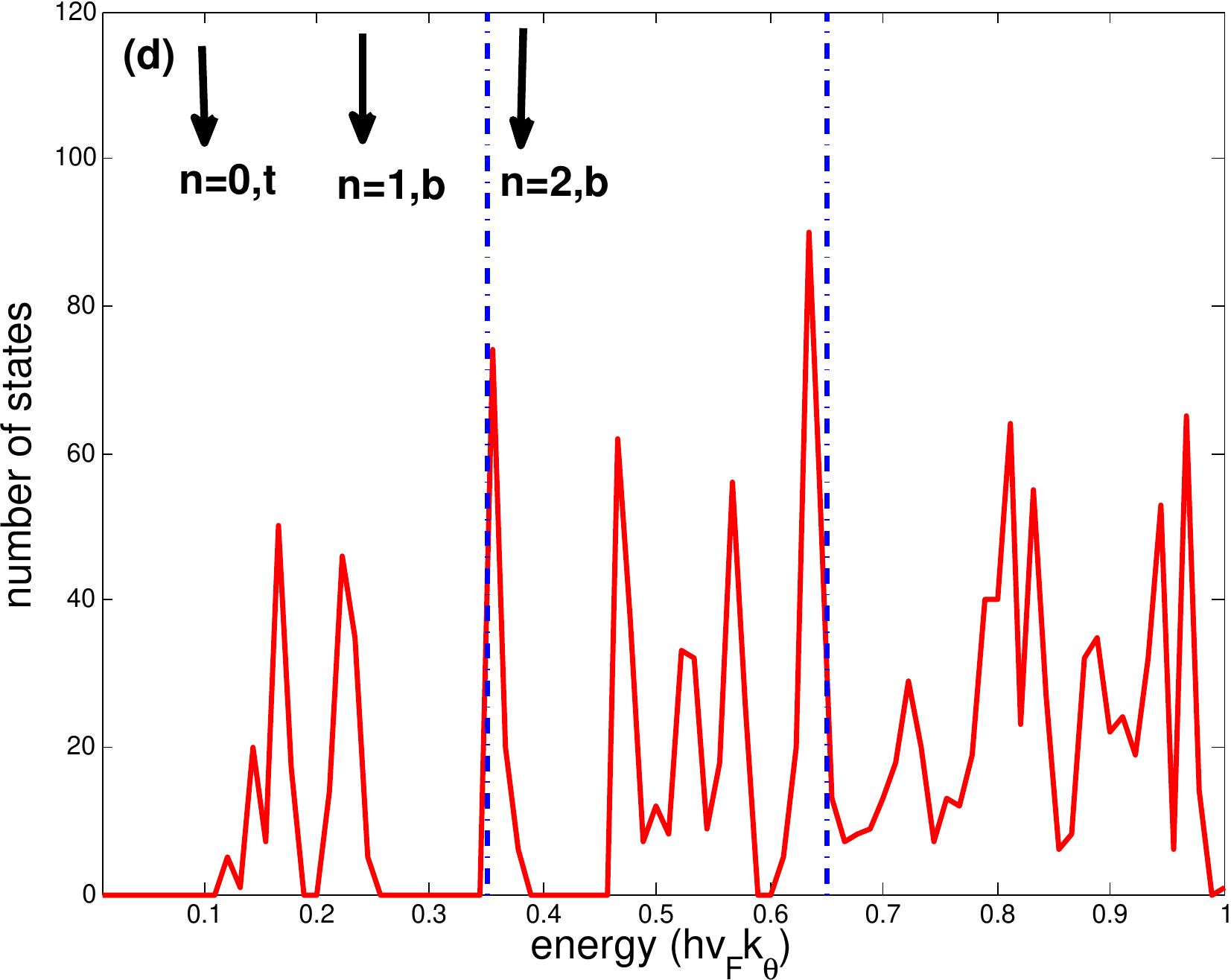}
\caption{(Color online). The density of states of twisted bilayer graphene with $\theta=5^{\circ}$ and $w=0.15$ are shown. In (a) and (c) [(b) and (d)] the interlayer bias $u=0$ ($u=0.2$), and the magnetic field is $B=$ 28.5 (38.8) Tesla in (a) and (b) [(c) and (d)]. Arrows are used to indicate the first and second Landau levels in the case without interlayer coupling. The saddle point $E_{sp}=1/2-w$ and  bottom of quadratic band $E_{bq}=1/2+w$ in the zero-field band structure are marked by the blue dashed lines.
}\label{dos}
\end{figure}
\end{widetext}

The spectrum of twisted bilayer graphene is obtained by diagonalizing the matrix 
representation of the Hamiltonian in Eq.\ (\ref{BasicModel}) in the basis
$\{|n,m\rangle,n\geq0, m\geq-n\}$, keeping only states consistent with
a particular choice of $\nu$ in Eq.~\ref{Phi_cond}. For convenience we adopt $\hbar v_Fk_{\theta}$ as our unit of energy.
The magnetic field in tesla ($T$) is then given by

\be
	B=\frac{8.5\times 10^4}{(k_{\theta}\ell)^2}\theta^2 [T]\:,
\ee
where the twist angle $\theta$, in units of radians in this equation,
is assumed to be small. $\ell\equiv \sqrt{\hbar c / eB}$ is the magnetic length.
For concreteness
we consider a twisted bilayer graphene with $\theta=5^{\circ}$, which corresponds to $\hbar v_F k_{\theta}=665$ meV. We also consider a relatively strong field regime, in the sense that the dimensionless number $k_{\theta}\ell$ is of order one. By doing so, we need only
a relatively small number of Landau levels to obtain results well-converged
with respect to the maximum Landau level index retained in the calculation. In addition,
we retain the same number of angular momentum states for each Landau level.  The results
we present are also well-converged with respect to this number.

Fig. \ref{dos} shows the density of states for the $\nu=0$ sector of the Hamiltonian in
Eq.\ (\ref{BasicModel}) for two magnetic fields, with and without interlayer bias.
(Results for $\nu=1$ and $\nu=2$ are almost identical to this, so that distinguishing among
the three sectors from their densities of states appears to be impractical.)
The panels (a) and (b) are for $B=28.5T$, and in (c) and (d) $B=38.8T$;
upper panels (a) and (c) are results without interlayer bias, while in the lower ones (c) and (d)
$u=0.2$.  The arrows indicate positions of Landau levels in the absence
of interlayer tunneling, and for the lowest few Landau indices $n$ we can see that
even with interlayer tunneling, density of states peaks can be
assigned a definite Landau level index, although they are broadened relative
to the case with no interlayer tunneling.  This broadening becomes more
pronounced with increasing $n$, and is especially pronounced
when the unperturbed level approaches the saddle point energy
$E_{sp}=1/2-w$~\cite{LuFertig14}.  An example of this can be seen in Fig.~\ref{dos}(a),
where the $n=2$ peak overlaps with $E_{sp}$ (shown as a dashed blue line in the figure.)
One may also see the $n=1$ peak is broader in (c) than in (a), where
in the former case it is closer to $E_{sp}$.
Therefore, we may identify $E_{sp}$ as an {energy scale} separating regimes
where the effects of interlayer tunneling are perturbative from one in which the
density of states is qualitatively affected by interlayer tunneling.
For energies above the quadratic band edge $E_{qb}=1/2+w$~\cite{LuFertig14},
and $u=0$, the density of states continues to oscillate but no longer actually
vanishes, indicating that Landau level mixing has become important at this
energy scale. Finally, we note also that the zeroth Landau level in our calculation
is broadened and even split by the interlayer coupling. This effect appears to be
particularly pronounced in strong fields, and can also be seen in
tight-binding calculations~\cite{Landgraf2013}.

The bottom two panels of Fig. \ref{dos} illustrate the effect of interlayer bias.
One may see that the Landau level peaks below $E_{sp}$ are shifted by $\pm u$,
but the features above this energy are less sensitive to bias, particularly
in the range  $E_{sp} < E < E_{qp}$.
This is clearest in panel (b), where the minima in the spectrum
above $E_{sp}$ move little in response to $u$, although the peak heights
and widths show some sensitivity to bias.  With the exception of the $n=2$
level associated with the bottom layer, this also seems to be the case in panel (d).  Above $E_{qb}$, some minima are relatively unaffected
by $u$, but some are, and at high enough energy the overall line shape
changes significantly when $u$ is introduced [most noticeably
in our results in panel (b)]

\begin{figure}
\input{epsf}
\includegraphics[width=0.5\textwidth]{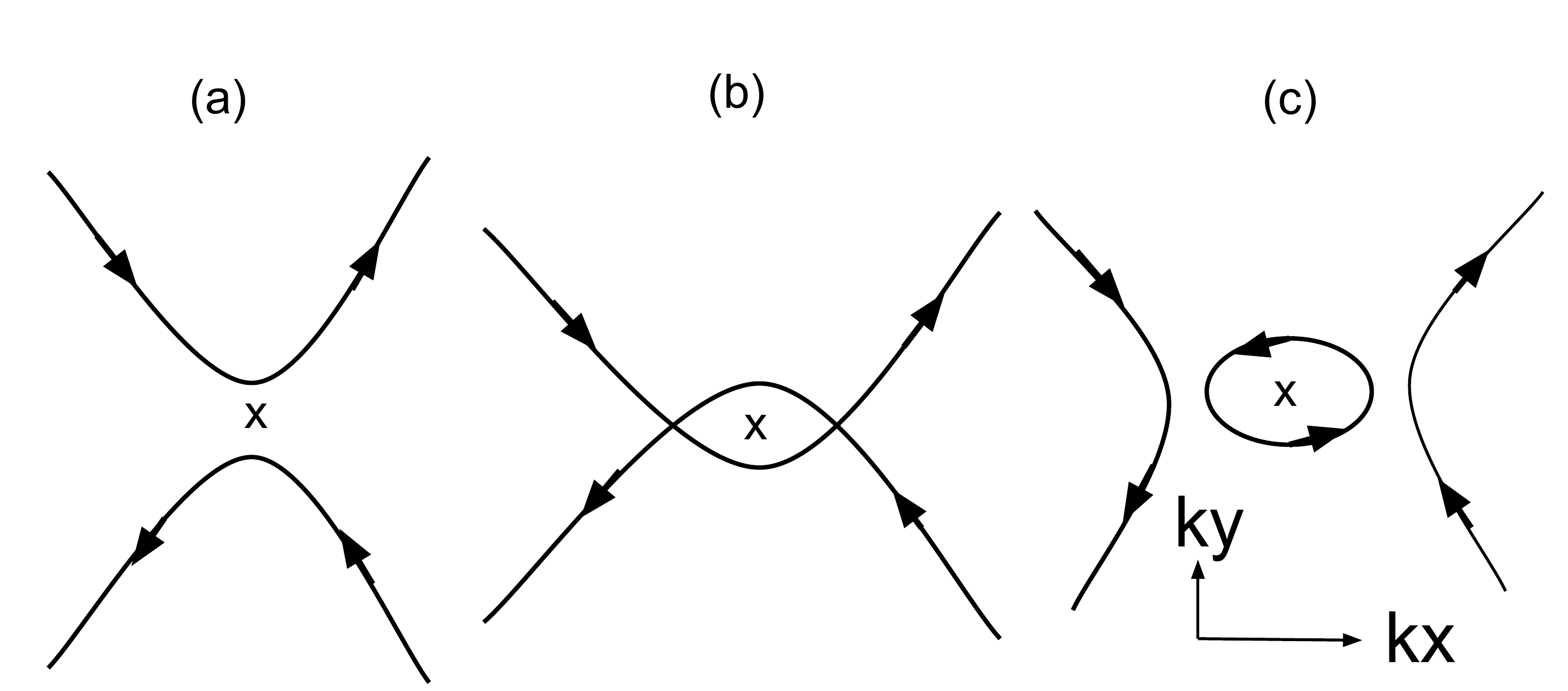}
\caption{ Evolution of semiclassical orbits in momentum space, near a saddle point. Panel (a) represents an orbit with energy below the saddle point with the upper (lower) part representing the trajectory in the top (bottom) layer.
The tunnel coupling between these orbits is relatively weak and can be well-approximated as localized
in a single layer. As the energy increases above $E_{sp}$, in the absence of interlayer coupling
the two orbits cross (b).  With tunneling there are two new orbits as illustrated in (c), and
the closed orbit illustrated represents a branch of allowed states when the energy exceeds
$E_{qb}$.  The reconnected orbits now oscillate between layers, and the corresponding states are manifestly delocalized between them.}\label{orbit}
\end{figure}

This mixed behavior may be understood in terms of the semiclassical description
of wavefunctions in a magnetic field, which are built up from semiclassical
orbits in the Brillouin zone, leading to
some some wavefunctions being more strongly coupled across the layers than others.
Fig.~\ref{orbit} illustrates one way in which this description leads
to such a change in character.  For states with energy well
below $E_{sp}$, the semiclassical orbits~\cite{LuFertig14} are weakly
tunnel-coupled between layers across saddle points
[Fig.~\ref{orbit}(a)], with this coupling increasing greatly as the energy rises
above $E_{sp}$ [Fig.~\ref{orbit}(c)].
When tunneling
can be ignored, wavefunctions can be associated with a particular layer,
and their energies will shift by $\pm u$ when interlayer bias is introduced.
When interlayer coupling is strong, one may consider the effect of $u$
within first order perturbation theory, and the energies of states that are equally
distributed between the two layers will be unaffected by bias to this order.
Thus, the sensitivity of the density of states to interlayer bias offers
a measure of how strongly coupled the wavefunctions are between the two
layers at a given energy.

We next turn to calculations of electromagnetic absorption (i.e., cyclotron resonance),
which which we will see is directly impacted by the behaviors residing in the density
of states.

\section{Cyclotron resonance}

In this section, we analyze electromagnetic absorption -- i.e., cyclotron resonance -- in this system,
to see what one learns about the energy spectrum and also the extent to which wavefunctions
are localized in one or the other layer.  We will consider the absorption of linearly polarized
waves, in which the polarization vector may be either parallel or perpendicular to the
plane of the sample.
Only the first of these induces transitions in single layer systems. In the bilayer system,
the second can induce transitions between states if one or both of them is significantly
delocalized between the layers.  In this way cyclotron resonance can give us
information about the nature of the wavefunctions.

\subsection{In-plane electric field}

\begin{widetext}

\begin{figure}
\input{epsf}
\includegraphics[width=0.4\textwidth]{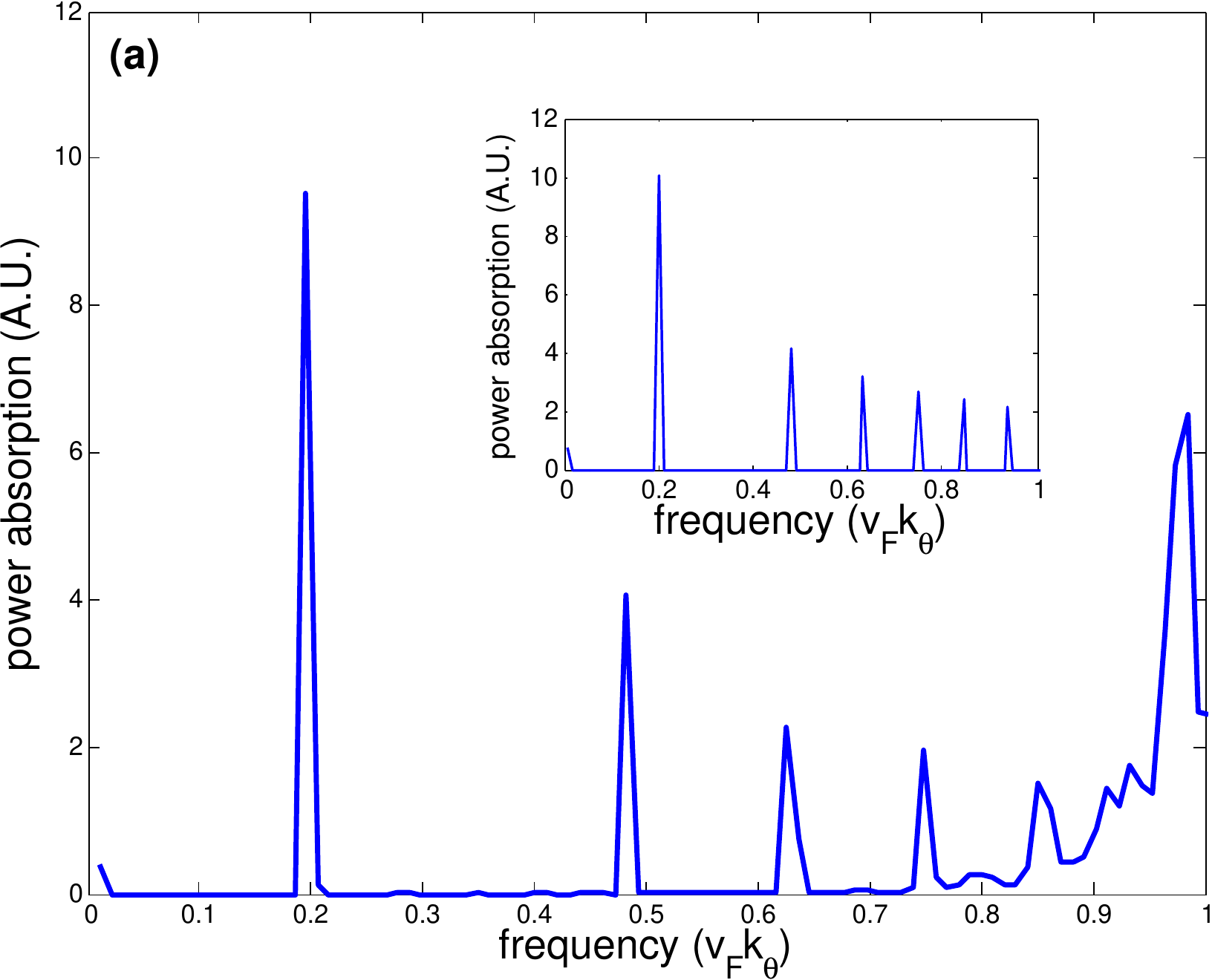}
\includegraphics[width=0.4\textwidth]{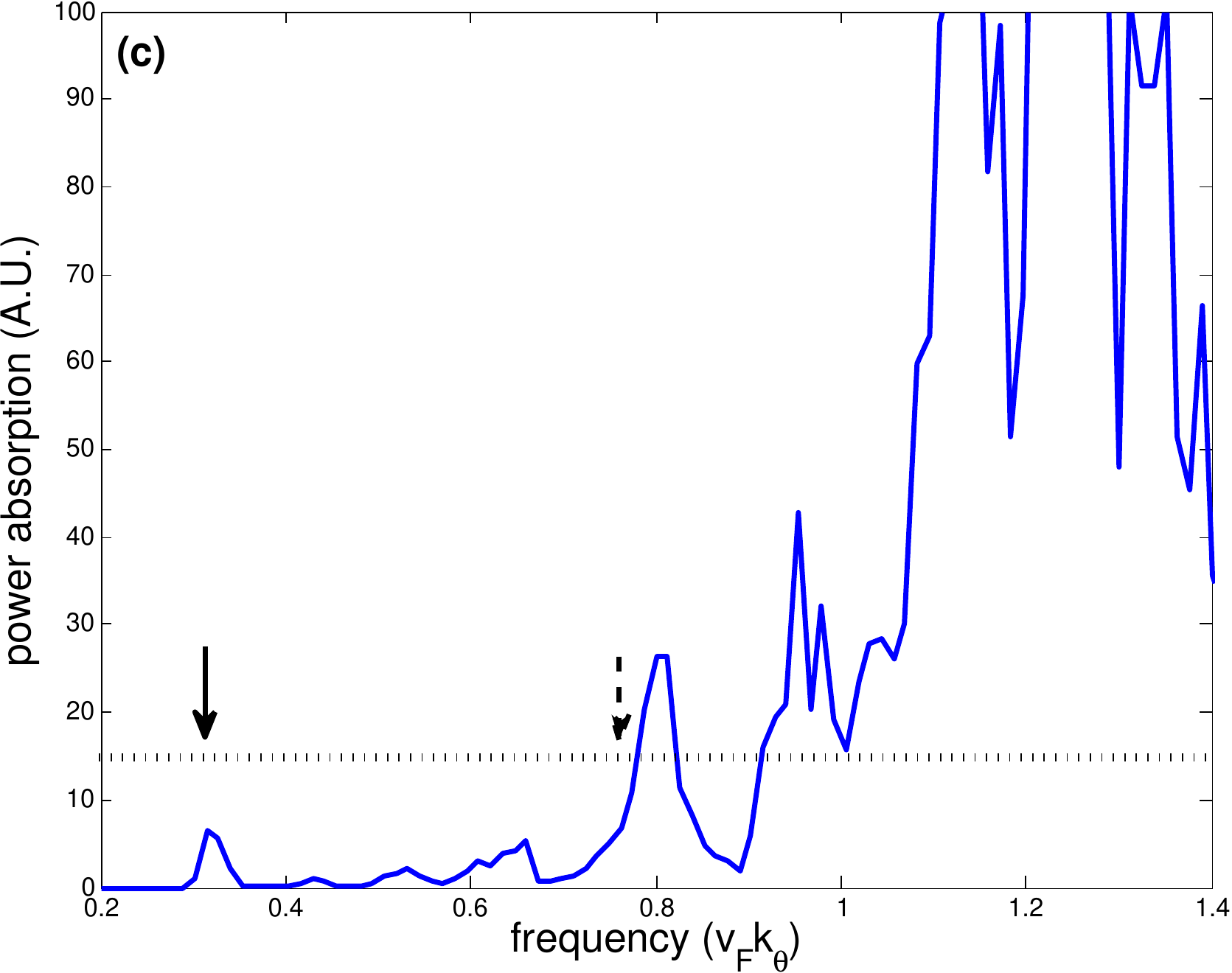}
\includegraphics[width=0.4\textwidth]{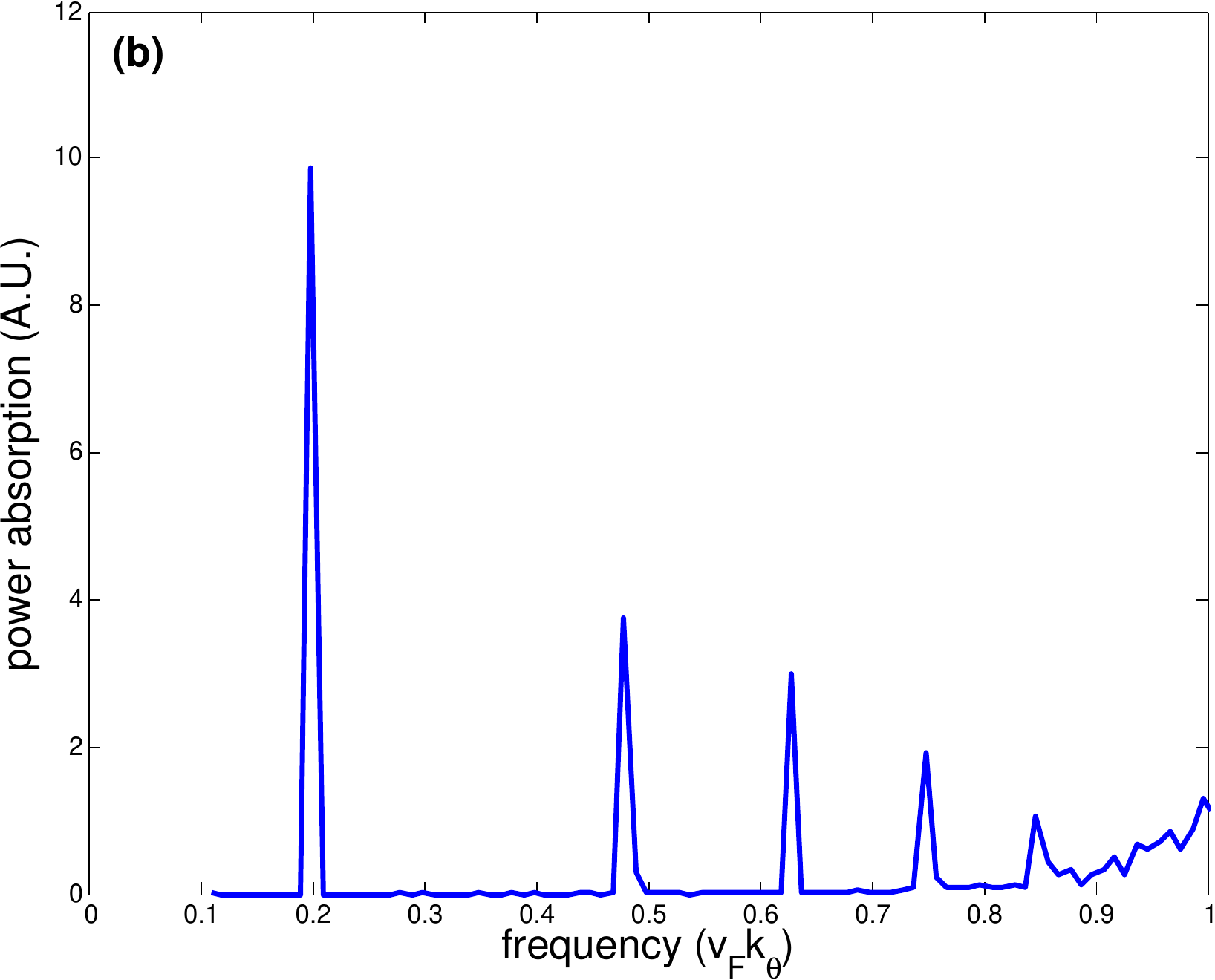}
\includegraphics[width=0.4\textwidth]{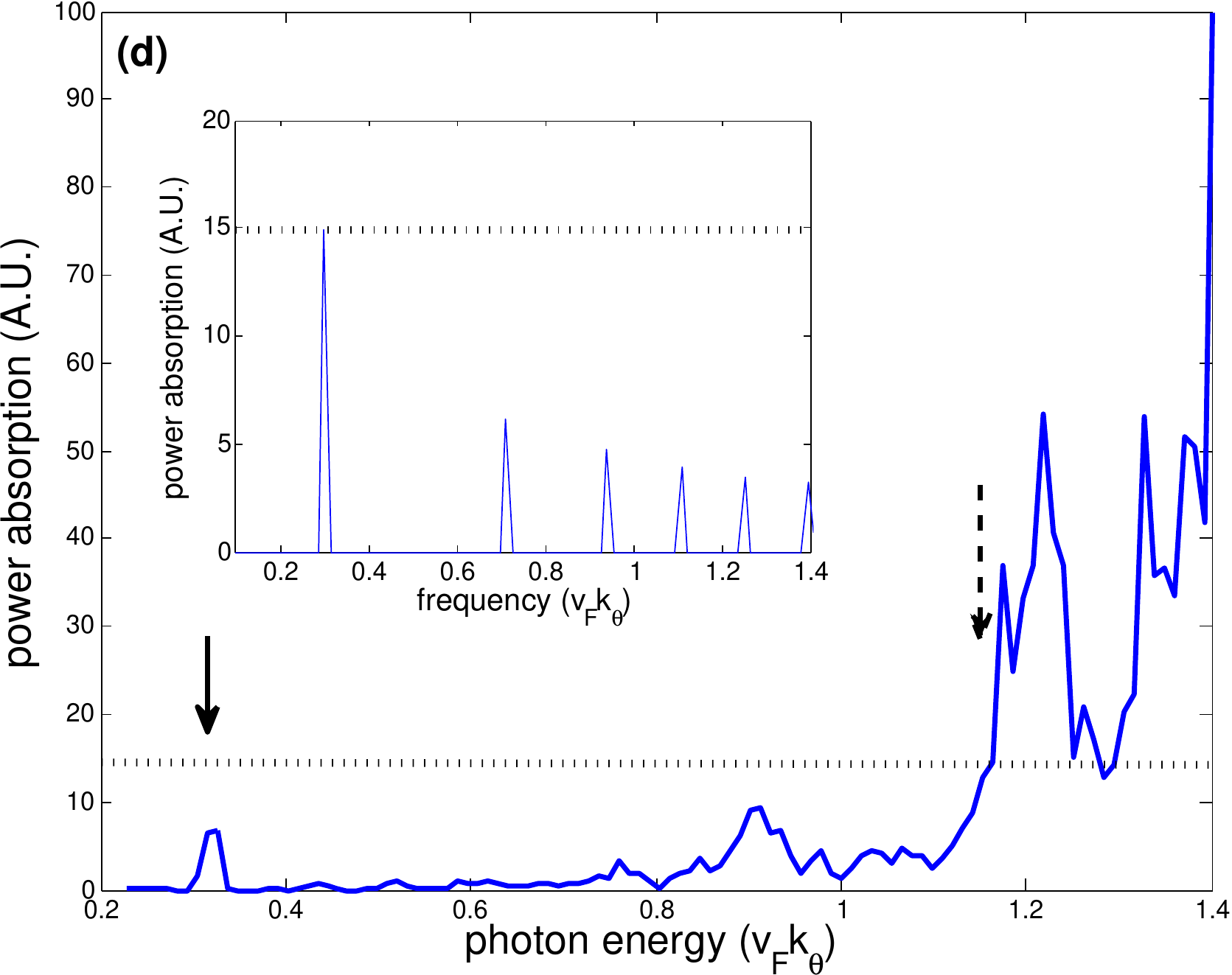}
\caption{(Color online). Results of the cyclotron resonance for $\mu=0$ with the external electric field lying in plane. The magnetic field $B$ = 13 (28.5) Tesla in panel (a) and (b) [(c) and (d)]. The top panels (a) and (c) have zero interlayer bias while the bottom panel (c) [(d)] has the interlayer bias $u=$ 0.1 (0.2). The insets in (a) and (d) shows the corresponding results in the absence of interlayer coupling.}\label{X150}
\end{figure}

\end{widetext}

When the electric field of the perturbing radiation is in-plane, the resulting perturbation,
with appropriate alignment with respect to the coordinate axes,
is proportional to

\be
	\hat x=\frac{a+a^{\dag}+b+b^{\dag}}{\sqrt{2}}\:,
\ee
where the magnetic length $\ell$ has been set to one.
The ladder operators in this expression acting on the basis states obey the relations

\bea
	a^{\dag}|n,m\rangle=\sqrt{n+1}|n+1,m-1\rangle\:,\\
	\ b^{\dag}|n,m\rangle=\sqrt{n+m+1}|n,m+1\rangle\:.
\eea
The operator $\hat x$ always changes the angular momentum by one, from which we can conclude that $\langle\Phi^{\nu}|\hat x|\Phi^{\nu}\rangle$ must vanish. Thus only transitions between
sectors of different $\nu$ can contribute to cyclotron resonance in this configuration.
From Fermi's Golden Rule,
for a given chemical potential $\mu$,
the energy absorption $\alpha$ as a function of frequency $\omega$
becomes

\be
	\alpha(\omega)/\alpha_0=\omega\sum_{\hbar\omega_f>\mu}\sum_{\hbar\omega_i<\mu}|\langle f|\hat x|i\rangle|^2\delta(\omega_f-\omega_i-\omega)\:,\label{power_X}
\ee
where $\alpha_0$ is an overall absorption scale, $|i(f)\rangle$ is the initial (final) single
particle state, and $\hbar\omega_{i(f)}$ is the energy of that state.
$\alpha(\omega)$ is essentially the real part of the dynamical conductivity \cite{Gusynin,Moon3},
and our matrix diagonalization gives us access to both the wavefunctions and energies
so we may compute it directly.

The results with chemical potential $\mu=0$ are illustrated in Fig.~\ref{X150}. For single
graphene layer, the corresponding absorption peaks satisfy a selection rule
$|n_f|=|n_i|\pm1$~\cite{Jiang07}; the resulting spectrum is illustrated in the insets of Fig.~\ref{X150}(a) and (d) for $B=$ 13 and 28.5 Tesla, respectively. If we label the frequency for the first peak associated with the transition $n=0\rightarrow n=1$ as $\omega_{01}$, then one can find the next three peaks at the frequencies of $(1+\sqrt{2})\omega_{01}$, $(\sqrt{2}+\sqrt{3})\omega_{01}$, and $(2+\sqrt{2})\omega_{01}$.
In a twisted bilayer, at low enough excitation frequencies only
eigenstates with $|E|<E_{sp}$ can be involved, and as discussed above these
states to a good approximation may be treated as layer-polarized. Thus we
expect to see the same absorption peaks as in single-layer graphene for sufficiently
low frequency. This expectation is indeed the case in our calculations. In panel (a), it can be seen that $\omega_{01}\approx0.2$ and the first four peaks in the single-layer case also appear at same frequencies in the tBLG. In panel (c) where the field is stronger, however, only the peak
indicated by solid arrows, at $\omega_{01}\approx0.3$ is found to coincide with its counterpart in single-layer case. This can be inferred from Fig. \ref{dos}(a) where the Landau levels of $n\geq2$ are significantly distorted by
interlayer coupling, and we begin to see deviations from the higher order
peaks in the single layer response. Those peaks representing the layer-polarized nature of states become broader with increasing field
once the final states move close to the saddle point, and eventually merge into the high-frequency spectra.

It is interesting to note that in the presence of interlayer bias --
for example, the spectrum with $u=0.2$ illustrated in Fig.~\ref{X150}(d) -- this peak is
essentially the same in position and oscillator strength as for $u=0$,
further confirming the single-layer nature of Landau levels with $|E|<E_{sp}$. For lower field, such insensitivity to $u$ is even more obvious as shown in panel (b) in which $u$=0.1 and the spectrum does not deviate from that in panel (a) below  $\omega \sim 0.9$.

At higher frequencies, $\alpha(\omega)$ deviates more dramatically from the single layer behavior.
Absorption is increased significantly when high-energy, layer-delocalized states participate in  transitions. We indicate a threshold energy for this change in behavior
by dashed arrows in Fig. \ref{X150}(c) and (d), which we define as the lowest frequency at
which the absorption exceeds the largest peak in the single-layer case (see inset).
Note that this threshold energy does depend on the interlayer bias, as illustrated in the figure.
We attribute the sensitivity to $u$ to the layer delocalized nature of eigenstates for $|E|>E_{sp}$. It is interesting to notice that the overall scale of absorption is much higher than
what is found in single layer electromagnetic absorption at these relatively low
frequencies.  Although $\alpha(\omega)$ obeys an oscillator strength sum rule \cite{Sabio},
this is possible because the absorption in the continuum model continues to
high frequency, with peak heights only falling off as $1/\omega$ at large frequency.
A finite value for the sum rule is only obtained
by imposing an energy cutoff in the spectrum.  Evidently, the interlayer coupling
transfers much of this high frequency oscillator strength for uncoupled layers to
much lower frequency for tBLG.  Clearly this is a non-perturbative effect.

Our general observation for this type of measurement is that at low frequency one
expects to reproduce the single layer absorption spectrum, with strong deviations
setting in once levels above $E_{sp}$ become involved.  In principle, by examining
the spectrum as a function of field and observing the frequency scale at which
the deviation sets in, one may infer the value of $E_{sp}$ for the system.
It is also possible to make this identification by looking
for very low-energy, intra-Landau level absorption, which first becomes significant
when a Landau level approximately coincides in energy with $E_{sp}$.  We discuss
this possibility in the next section.

\subsection{Out-of-plane electric field}

\begin{figure}
\input{epsf}
\includegraphics[width=0.3\textwidth]{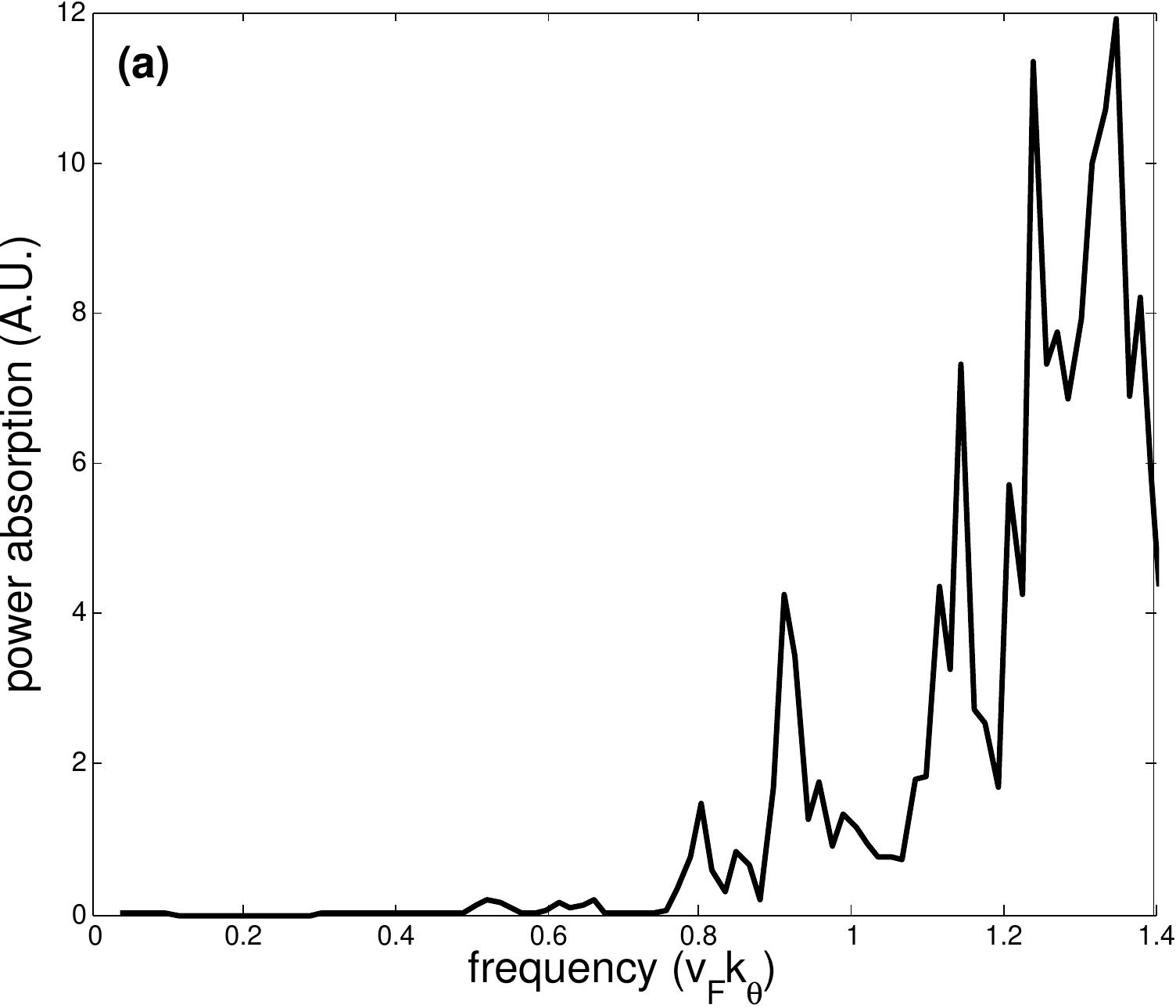}
\includegraphics[width=0.3\textwidth]{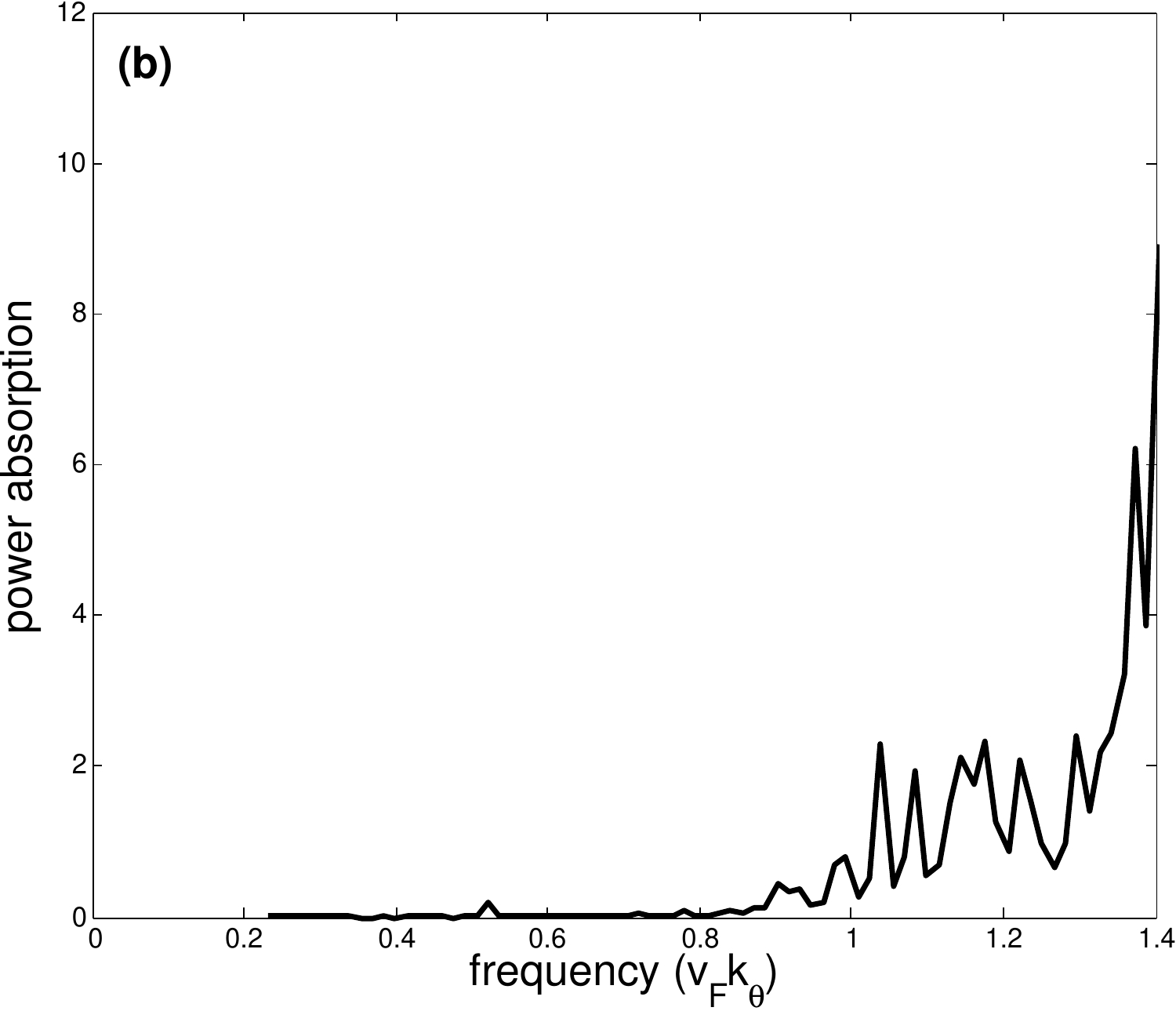}
\caption{Results of the cyclotron resonance for $\mu=0$ with the external electric field parallel with the normal of plane. The magnetic field $B$ = 28.5 Tesla, and panel (a) and (b) are results without and with interlayer bias.}\label{Z150}
\end{figure}

Next we consider the electric field polarized along the normal to the graphene plane.
As remarked above, this can induce transitions because of the bilayer structure.
In this case, the perturbation due to the vertical electric field is, up to an overall constant, the Pauli matrix $\tau_z$,
acting on the two-dimensional layer space.
This preserves the symmetry of the rotation operator $G$ in Eq.\ (\ref{G_sym}).
Thus only the transitions within same sector $\nu$ are possible.  The absorption
as a function of frequency in this case can be written as

\be
	\alpha_V(E)/\alpha_V^0=\omega\sum_{\hbar\omega_f>\mu}\sum_{\hbar\omega_i<\mu}|\langle f|\tau_z|i\rangle|^2\delta(\omega_f-\omega_i-\omega)\:.
\ee

Fig.~\ref{Z150}(a) and (b) show the results for the out-of-plane configuration with and without interlayer bias. In contrast with the in-plane configuration, it can be seen that the transition associated with single-layer behavior is completely absent, so that features relatively independent of interlayer bias $u$ are not present.
The threshold frequency beyond which absorption increases markedly is sensitive to the interlayer bias,
just as in the in-plane polarization case.

We may also use this configuration to explore the saddle point in the band structure. As
mentioned above, the power absorption will vanish in the limit that the interlayer
coupling $w \rightarrow 0$, so that absorption in this configuration is sensitive
to whether the states are delocalized across the layers. One way in which this is
manifested is the broadening of Landau levels as they cross the saddle point with
increasing field.
This admits the possibility of intra-Landau level absorption.
Such absorption is highly suppressed in single layers because it can only occur
when there is some perturbation that can relax the selection rule.  In this case
the perturbation is the interlayer coupling.

\begin{widetext}

\begin{figure}
\input{epsf}
\includegraphics[width=0.4\textwidth]{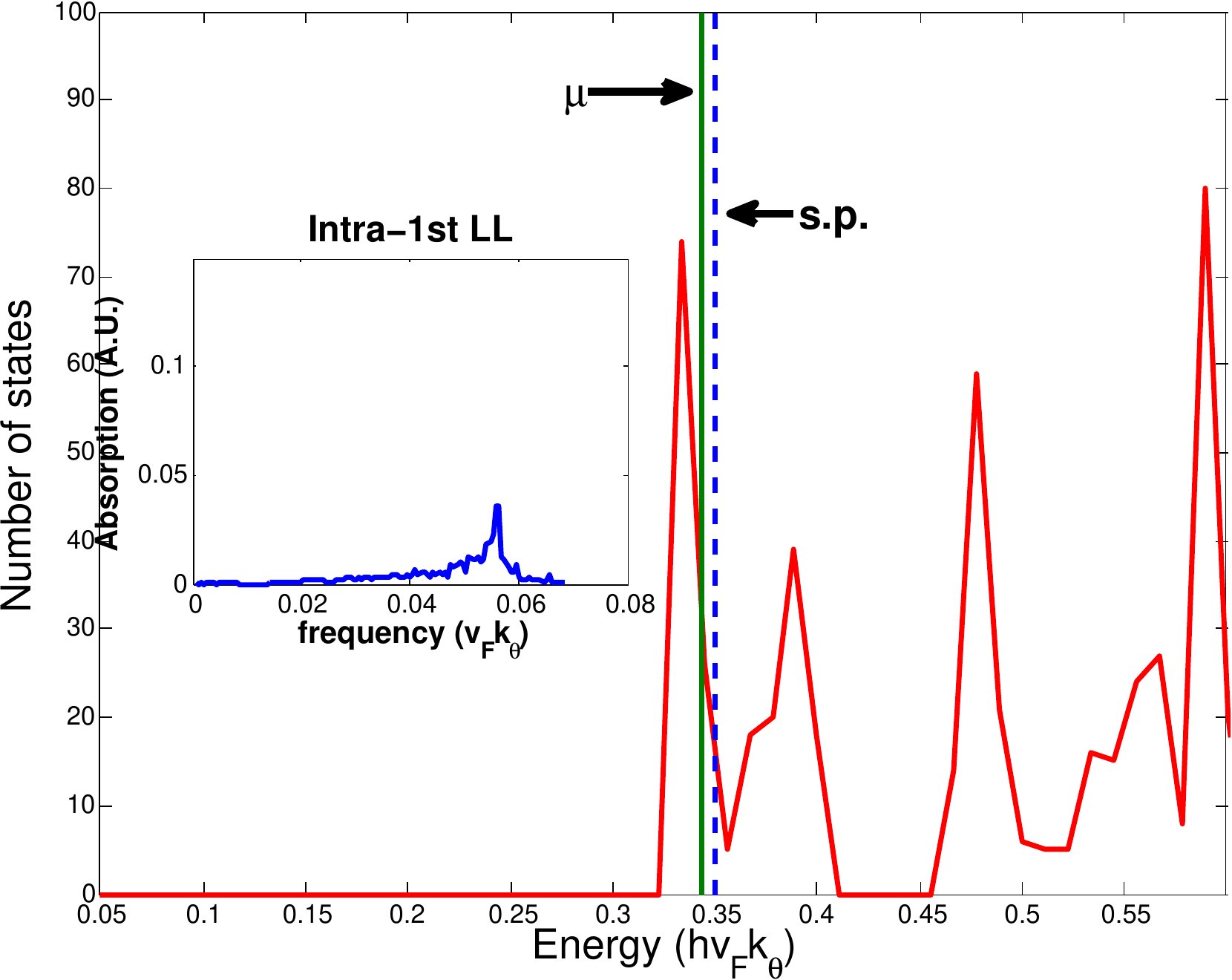}
\includegraphics[width=0.4\textwidth]{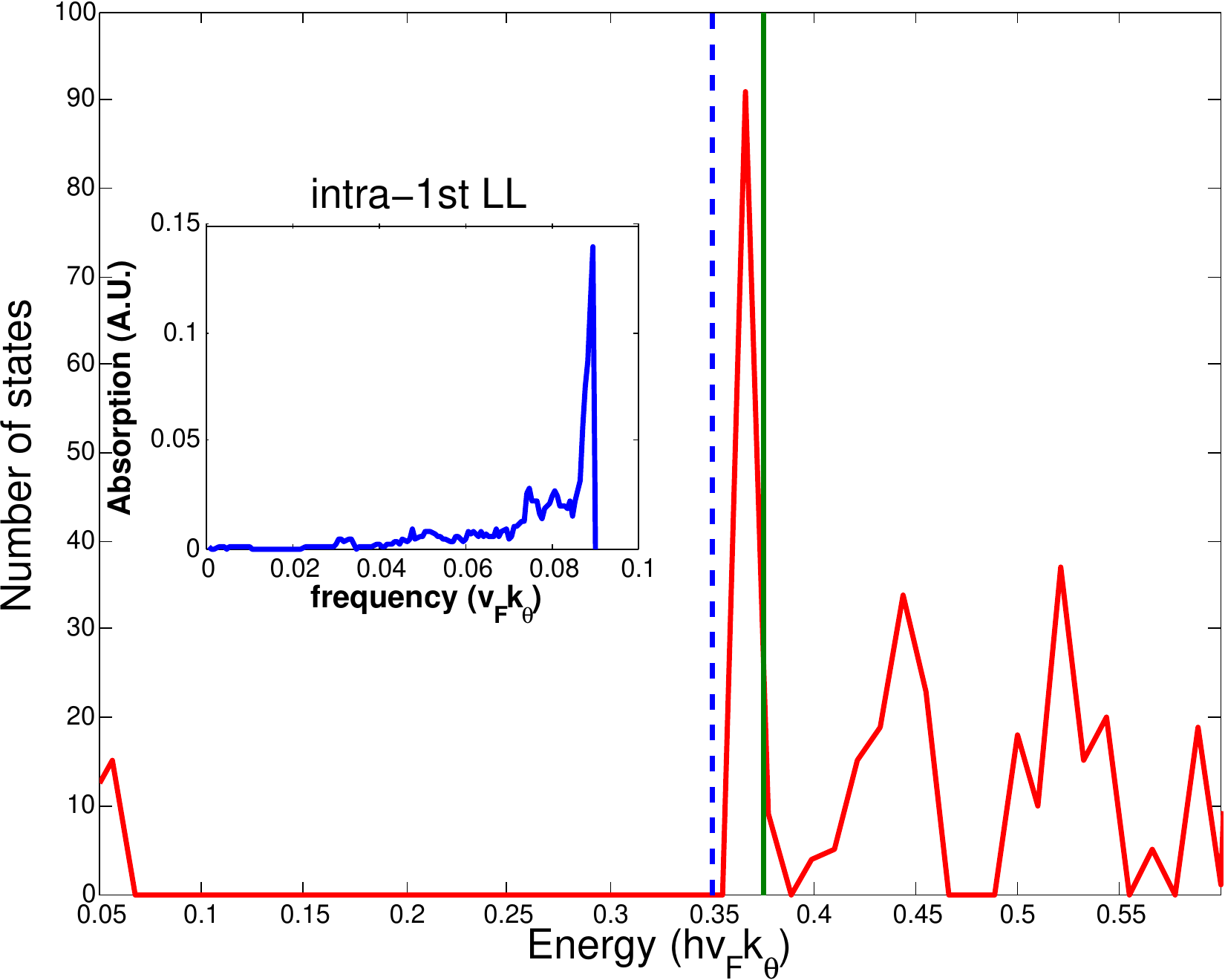}
\caption{(Color online). Evolution of the density of states as the first Landau level passes through the saddle point
(marked by the blue dashed line) with increasing field. In the right (left) panel, $B=$ 38.8 (46.6) Tesla. In the insets,
the power absorption resulting from excitations within the first Landau level
are shown. (Fermi levels are marked by green solid lines in the main panels.)
A significant increase is seen as the first Landau level passes through the saddle point.}\label{evolveLL}
\end{figure}

\end{widetext}

\begin{figure}
\input{epsf}
\includegraphics[width=0.4\textwidth]{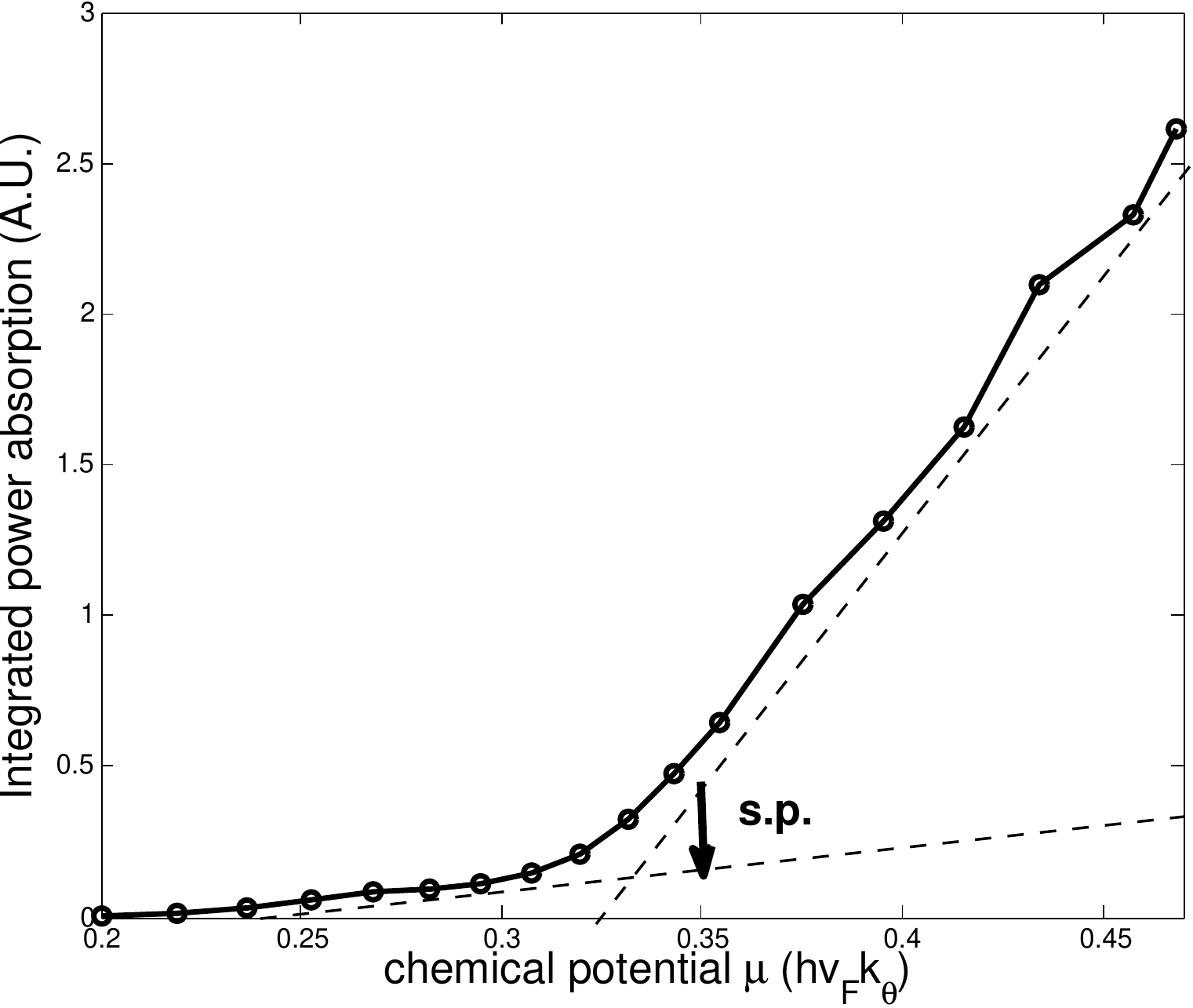}
\caption{Integrated power absorption for the intra-Landau level transition. Horizontal axis: the chemical potential which is tuned to coincide with the center of first Landau level as the magnetic field varies. Vertical axis: the integrated power absorption due to the transition within the first Landau level. The two asymptotic lines intersect at an 
energy close to the saddle point.}\label{power_sp}
\end{figure}

To carry this out one must be able to locate the Fermi level so that half the states of the broadened Landau level are filled, which in principle is possible via gating.  One then adjusts the magnetic field so that the first Landau level passes through the saddle point (while
keeping the level half filled.) Fig.\ \ref{evolveLL} shows two density of states
plots in which the $n=1$ Landau levels are pushed toward higher energy, broadening and
and deforming the peak in the process. This broadening creates the possibility of electromagnetic
absorption at very low-frequencies
due to the intra-Landau level transitions.
Shown in the insets are the absorption spectra, which now contain
a low-energy peak.  Comparing the two peaks for the Landau level just below and just above
the saddle point, one can see a dramatic increase in both the maximum peak height and
its oscillator strength as the Landau level center passes from just below to just above
the saddle point energy.
Fig.\ \ref{power_sp} shows results for the oscillator strength (obtained by numerically
integrating the area under the peaks such as shown in the insets of Fig.\ \ref{evolveLL})
as a function of the ``center'' of the Landau level.  This oscillator strength essentially
starts to grow for $\mu$ just below $E_{sp}$, and has a sharp increase
close to, although slightly below, the actual value of $E_{sp}$.
In principle this provides a way to locate
the energy of the saddle points separating the two Dirac points of the twisted bilayer.

\section{Conclusions}

We have studied the density of states and cyclotron resonance for
Landau levels of tBLG within a continuum model.  The model supports a three-fold symmetry
which allows the states to be separated into three sectors, allowing some
simplification of the numerical calculations.
We find that the spectrum in a magnetic field evolves from discrete levels closely akin
to single layer Landau levels
into increasingly overlapping bands with increasing energy.
Below the saddle point, the ``band width" is very small and the magnetic states are mainly layer-polarized.  Above the saddle point, the magnetic states becomes strongly layer-delocalized
and the bands are significantly broader. This behavior follows from semiclassical descriptions
of the wavefunctions, in which orbits in different layers begin to cross one another
above the saddle point energy.

In order to explore the impact of interlayer coupling and
related saddle points in the tBLG band structure,
we propose two configurations for measuring
electromagnetic absorption (cyclotron resonance) in this system. In one configuration
the external electric field is polarized along the in-plane direction,
and we find that as long as the magnetic field is not too strong, states below the saddle point are effectively layer-polarized
and give rise to the same response as a single-layer system. On the other hand,
the responses at higher frequencies involve layer-delocalized states.  In the other configuration,
where the electric field is polarized perpendicular to the system,
layer delocalized states must be involved in the transitions.  The different kinds
of states may be distinguished by whether they can contribute to both types of
absorptions, and by their sensitivity to interlayer bias.

We also found that the Landau level bands broaden significantly  above the saddle point,
so that power absorption becomes possible at very low frequency which is completely
absent in the single layer case.
In the perpendicular polarization configuration, we find the absorption increases significantly when the Landau band is elevated above the saddle point. In principle, this allows one to
locate the saddle point energy from the absorption spectra of tBLG.

{\it Acknowledgements} This work was supported in part by the NSF through Grant No. DMR-1005035,
and by the  US-Israel Binational Science Foundation.

\newpage


\begin{thebibliography}{plain}
\bibitem{comment} Strictly speaking, any eigenstate of the system will be delocalized between
both layers.  However, in general one may form linear combinations of states that are
localized in a single layer.  If these are close in energy then the state can be usefully thought
of as localized in a single layer.  This turns out to be especially true for states
at energies well below that of the saddle point.  Sensitivity of the wavefunction energies
to interlayer bias gives one measure of interlayer coupling inherent to the wavefunctions.
\end{thebibliography}
\end{document}